\documentclass[11pt,reprint,aps,prb,amsmath,amssymb,showpacs,superscriptaddress]{revtex4-1}
\pdfoutput=1

\usepackage{graphicx}
\usepackage{dcolumn}
\usepackage{bm}

\usepackage{xspace}

\def\phm#1{\phantom{#1}}
\def\beq{\begin{equation}}
\def\eeq{\end{equation}}
\def\rb{{\bf r\xspace}}
\def\R{{\bf R\xspace}}
\def\k{{\bf k\xspace}}

\newcommand{\EF}{$E_{\mathrm{F}}$\xspace}
\newcommand{\ket}[1]{|#1\rangle}
\newcommand{\bra}[1]{\langle#1|}
\newcommand{\braket}[2]{\langle#1|#2\rangle}

\begin{document}

\title{
Imaging the buried MgO/Ag interface: formation mechanism of the STM contrast
}

\author{Andrei Malashevich}
\email{andrei.malashevich@yale.edu}
\affiliation{Center for Research on Interface Structures and Phenomena (CRISP),
Yale University, New Haven, Connecticut 06520, USA}
\affiliation{
Department of Applied Physics, Yale University,
New Haven, Connecticut 06520, USA }
\author{Eric I. Altman}
\affiliation{Center for Research on Interface Structures and Phenomena (CRISP),
Yale University, New Haven, Connecticut 06520, USA}
\affiliation{Department of Chemical and Environmental Engineering,
Yale University, New Haven, Connecticut 06520, USA}
\author{Sohrab Ismail-Beigi}
\affiliation{Center for Research on Interface Structures and Phenomena (CRISP),
Yale University, New Haven, Connecticut 06520, USA}
\affiliation{
Department of Applied Physics, Yale University,
New Haven, Connecticut 06520, USA }
\affiliation{
Department of Physics, Yale University,
New Haven, Connecticut 06520, USA }
\affiliation{
Department of Mechanical Engineering and Materials Science, Yale University,
New Haven, Connecticut 06520, USA }


\begin{abstract}
Scanning tunneling microscopy (STM) provides real-space electronic state information at the atomic scale that is most commonly used to study materials surfaces.  An intriguing extension of the method is attempt to study the electronic structure at an insulator/conductor interface by performing low-bias imaging above the surface of an ultrathin insulating layer on the conducting substrate.  We use first-principles theory to examine the physical mechanisms giving rise to the formation of low-bias STM images in the MgO/Ag system. We show that the main features of the low-bias STM contrast are completely determined by the atoms on the surface of MgO.  Hence, the low-bias contrast is formed by states at the Fermi level in the Ag that propagate evanescently through the lattice and atomic orbitals of the MgO on their way to the surface.  We develop a number of analysis techniques based on an {\em ab initio} tight-binding representation that allow identification of the origin of the STM contrast in cases where previous approaches have proven ambiguous.

\end{abstract}

\maketitle

\section{Introduction}
\label{sec:intro}

Metal oxide surfaces and interfaces involving metal oxides have been, and continue to be, a subject of significant scientific and technological interest due to the fact that metal oxides display a wide range of physical basic science phenomena that are simultaneously useful in technological applications.  For example, oxide surfaces are used for catalysis or as gas sensors,\cite{jackson_09,comini_09} oxide insulators are ubiquitous as gate insulators in transistors, and recent advances in layer-by-layer growth have triggered great activity in the study and design of oxide heterostructures.\cite{zubko_11,tsymbal_12}  In addition, the physical and chemical properties of metal oxides in thin film form can differ substantially from those of their bulk forms, as these properties are affected strongly by the presence of surfaces and interfaces and especially the substrate.  A classical example is provided by ultrathin films of an oxide such as MgO on a metallic substrate where electron transfer from the metal to the surface of the oxide can drive oxidation and other reactions.\cite{pacchioni_13} 

From the viewpoint of direct characterization of oxide thin films in real space, scanning tunneling microscopy (STM) is readily able to visualize the surface of a system.  A more intriguing possibility is to attempt to use STM on the surface of an oxide thin film to learn about the buried interface below the surface.  If one can selectively probe the interface versus the surface by, e.g., tuning the bias voltage, one has on hand a powerful method to study interfaces at the atomic scale and in real space without the requirement of long-range order typical of diffraction based methods.\cite{robinson_92,hashizume_92,specht_93,renaud_99}

One of the prototypical metal/oxide interfaces that has been studied extensively using STM in the past two decades is the $M$/MgO interface, where $M$ is either Mo or Ag. Let us briefly summarize the key findings about this system.
Gallagher \textit{et al.}\cite{gallagher_95} performed a 
scanning tunneling microscopy (STM) study on a thin MgO film grown on the
(001) surface of Mo and were able to obtain STM images for MgO thicknesses up to 25~\AA\ despite the insulating nature of MgO. Schintke \textit{et al.}\cite{schintke_01} 
performed scanning tunneling spectroscopy (STS) as well as first-principles calculations on a thin MgO film on Ag (001) and found that the STS spectrum was essentially flat for a range of bias voltages between $-4$ and $+1.7$ V; the first peak in the STS spectrum at $1.7$ V for the thinnest films disappeared with MgO thickness whereas a peak at $2.5$ V remained constant with thickness.  The thickness independence permitted an identification of the 2.5 V feature with electronic states belonging to the MgO which yielded an estimated band gap of 6.5 V for MgO.  The density functional theory (DFT) calculations of the local density of states (LDOS) for MgO thicknesses between 0 and 3 atomic layers in the same work found a good agreement with their experimental results.  Their main conclusion was that for bias voltages within the MgO band gap, the STM is probing Ag states through the MgO layer and that ultrathin MgO is required since the Ag wave functions decay exponentially in the insulating regions.

Lopez and Valeri\cite{lopez_04} performed
DFT calculations of the MgO/Ag (001) system
for one and two layers of MgO including several calculations with missing
oxygen atoms. They calculated STM images within the Tersoff-Hamann 
approximation.\cite{tersoff_83} Their main conclusions were in agreement
with the findings of Schintke \textit{et al.}\cite{schintke_01}: at low bias voltages, the
Ag states at the interface are probed whereas at high positive bias 
voltage the surface of MgO is probed.  Within this pictures, if an oxygen atom
is missing at the MgO/Ag interface, one would expect to see a bright STM feature
on top of the vacancy at low voltages from the Ag atom just below the vacancy, but such a bright feature was not actually seen, a difference attributed to geometrical effects due to the relaxation of the top MgO layer.

We note that in the case of the MgO/Ag (001) system, the MgO film is commensurate with the Ag substrate so that Mg and O atoms are in vertical registry with the Ag atoms along the (001) direction. Furthermore, the rock-salt nature of MgO insures that cation and anion atoms in consecutive (001) atomic layers switch identity.  In these conditions, assignment of observed STM features is ambiguous in theory and experiment.  A typical tool used theoretically is an analysis of the projected density of states (PDOS) onto the constituent atoms.  The PDOS is very useful as it directly identifies which atoms can contribute to tunneling at a given bias voltage, but unfortunately mere existence of a nonzero PDOS on some atom at a certain energy does not tell us to what extent the atomic orbitals of that atom actually contribute to the STM tunneling signal above the surface.  Hence, we believe that a more detailed theoretical investigation of this model system is necessary to determine precisely what electronic states are actually being observed in the surface STM measurement.

In this work, we have used a variety of first-principles calculations and analyses to study the origin and formation of the STM signal above the surface.  All results show that the STM signal at low bias (\textit{i.e.}, at the Fermi level) is completely dominated by the contribution of the atomic orbitals of the topmost MgO surface layer; atomic orbitals belonging to Ag atoms in the substrate make a negligible contribution.  Nevertheless, since the MgO film is insulating, the states at the Fermi level that are being imaged begin in the Ag substrate, couple to the atomic orbitals of the MgO in the interfacial region, and decay evanescently through the film on their way to the surface: the coupling is best understood as being through the lattice of the MgO and its associated atomic orbitals.  Unfortunately, the nature of the Ag-MgO coupling turns out to be complex and consists of interfering paths via both valence and conduction band states of the MgO film.

The body of the paper is organized as follows. Section \ref{sec:comp}
describes the details of our numerical simulations. Then the main results
are presented in Sec.~\ref{sec:results}. The results are further analyzed 
using a simplified tight-binding model is Sec.~\ref{sec:model}. 
The key findings are summarized in Sec.\ref{sec:summary} where we also provide our outlook for future work in this general area.

\section{Computational methods}
\label{sec:comp}

Our first-principles calculations use DFT 
within the generalized-gradient approximation (GGA)
of Perdew and Wang (PW91).\cite{perdew_91} We employ a planewave pseudo\-potential 
approach utilizing Vanderbilt ultrasoft pseudo\-potentials\cite{vanderbilt_90} as implemented in the {\sc Quantum ESPRESSO} software package.\cite{QE_09}
The atomic configurations and cutoff radii used for 
pseudopotential generation are shown in Table~\ref{tab:pseudo}.
For Ag and Mg atoms, non-linear core corrections are employed.\cite{louie_82}
The planewave kinetic energy and charge density cutoffs are 35 Ry and 
280 Ry, respectively. Supercells in the form of two-dimensional slabs are constructed with 
3 monolayers (ML) of face-centered cubic Ag with (001) surface orientation
and various thicknesses of rock-salt MgO.  Below, we label the MgO thickness by the number of ML (atomic planes) comprising the film.  The supercells contain a vacuum region of at least 20 \AA\
that separates each slab from its periodic images.
In addition, to ensure the absence of long-ranged electrostatic interactions
between the periodic slabs,
we use the dipole correction technique\cite{bengtsson_99}
to eliminate any electric field in the vacuum region entirely.

For a primitive interfacial unit cell that simulates MgO epitaxial to the Ag substrate (Fig.~\ref{fig:slab}), a $12\times12$ $k$-point sampling of the 2-dimensional Brillouin zone is employed.  Equivalent meshes of $k$ points are used for larger supercells.
For STM simulations,  we use the Tersoff-Hamann approximation.\cite{tersoff_83}
The main focus of the present work is on low-bias STM analysis for which Tersoff-Hamann approximation is an appropriate choice.  Specifically, since at low bias the STM current is proportional to the LDOS at the Fermi level \EF, we will be using the theoretically calculated LDOS as a surrogate for the STM signal, and will use the two terms interchangeably.

In Section \ref{sec:model} we will be constructing a tight-binding description of the MgO/Ag system.  The atomic orbitals comprising the tight-binding basis are those included in the pseudopotential generation and are listed in Table~\ref{tab:pseudo}.  In this basis, we compute matrix elements of the Kohn-Sham Hamiltonian as well as the atomic orbital overlap matrices.
This starting basis is normalized but not orthogonal (orbitals on different atoms have non-zero overlap).  In certain parts of the analysis in Section \ref{sec:model}, it is helpful to have an orthonormal atomic orbital basis: we use the L\"owdin transformation to arrive at an equivalent set of orthonormal L\"owdin orbitals.\cite{lowdin_50}  Finally, all projected density of states (PDOS) in this work employ L\"owdin orbitals for the projections.

\begin{table}
  \caption{\label{tab:pseudo}
  Pseudopotential reference valence configurations and corresponding
cutoff radii (atomic units).}
\begin{ruledtabular}
\begin{tabular}{lllll}
  Atom  & Valence configuration & $r_{\mathrm{c}}^s$ & $r_{\mathrm{c}}^p$ & $r_{\mathrm{c}}^d$ \\
\hline
Ag\phantom{$\int_a^{\int_a^b}$} & $4d^{10}5s^15p^0$ & 2.5 & 2.5 & 2.1  \\
                             Mg & $2p^63s^13p^{0.75}3d^0$ & 2.3 & 2.0 & 2.0 \\
                             O  & $2s^22p^6$ & 1.2 & 1.2 & \\
\end{tabular}
\end{ruledtabular}
\end{table}

\section{Results}
\label{sec:results}

\subsection{Bulk Ag and MgO}
\label{sec:bulk}

Our computed lattice constants of bulk fcc Ag and rock-salt MgO are $a_{\mathrm{Ag}}=4.13$~\AA\ and $a_{\mathrm{MgO}}=4.26$~\AA\, in good agreement with the respective experimental values\cite{wyckoff_63} of $4.09$~\AA\ and $4.21$~\AA.
Since these lattice parameters match to within $3\%$, epitaxial growth of MgO thin films on Ag (001) is feasible.\cite{schintke_01}

For MgO, our computed Kohn-Sham energy gap is $4.4$ eV which is much smaller than the experimental value of $7.9$~eV.\cite{williams_67}: the band gap underestimation of DFT is a well-known limitation of the theory.\cite{perdew_82,perdew_83}  However, this problem is not crucial for our study since the focus is on the low bias range about the Fermi level, we only need to insure that the MgO is an insulator whose band edges are well separated from the Fermi level, which is the case as shown below.  For reference, we show the density of states (DOS) and atomic PDOS for bulk MgO in Fig.~\ref{fig:pdos_mgo}(a).  As expected for an ionic oxide, the Figure shows that the valence band edge is dominated by anionic O $2p$ states while the conduction band is primarily a mixture of cationic Mg $3s$, $3p$, and $3d$ states.

At the lowest approximation level, STM probes the states between 
the Fermi level and the bias voltage with more weight on the upper
side of the energy range.\cite{chen_08}
For insulating MgO, the natural expectation is that for low bias 
voltages no  electronic states of the MgO itself will be probed since the Fermi level is solidly in the band gap.  Any STM image on the MgO surface must therefore originate in some way from the Ag substrate buried under the MgO film.  On the other hand, for large bias voltages (positive or negative), the Fermi level will enter the energy bands of the MgO film and a large signal should be observed coming from the MgO surface itself:  for large positive bias within the conduction band, Mg atoms should dominate as bright spots in the STM.  Conversely, for large negative bias in the valence band, the surface O atoms should dominate the STM image.

\subsection{2 ML MgO on 3 ML Ag(001)}
\label{sec:slabs}

\begin{figure}
\centering\includegraphics[width=4cm]{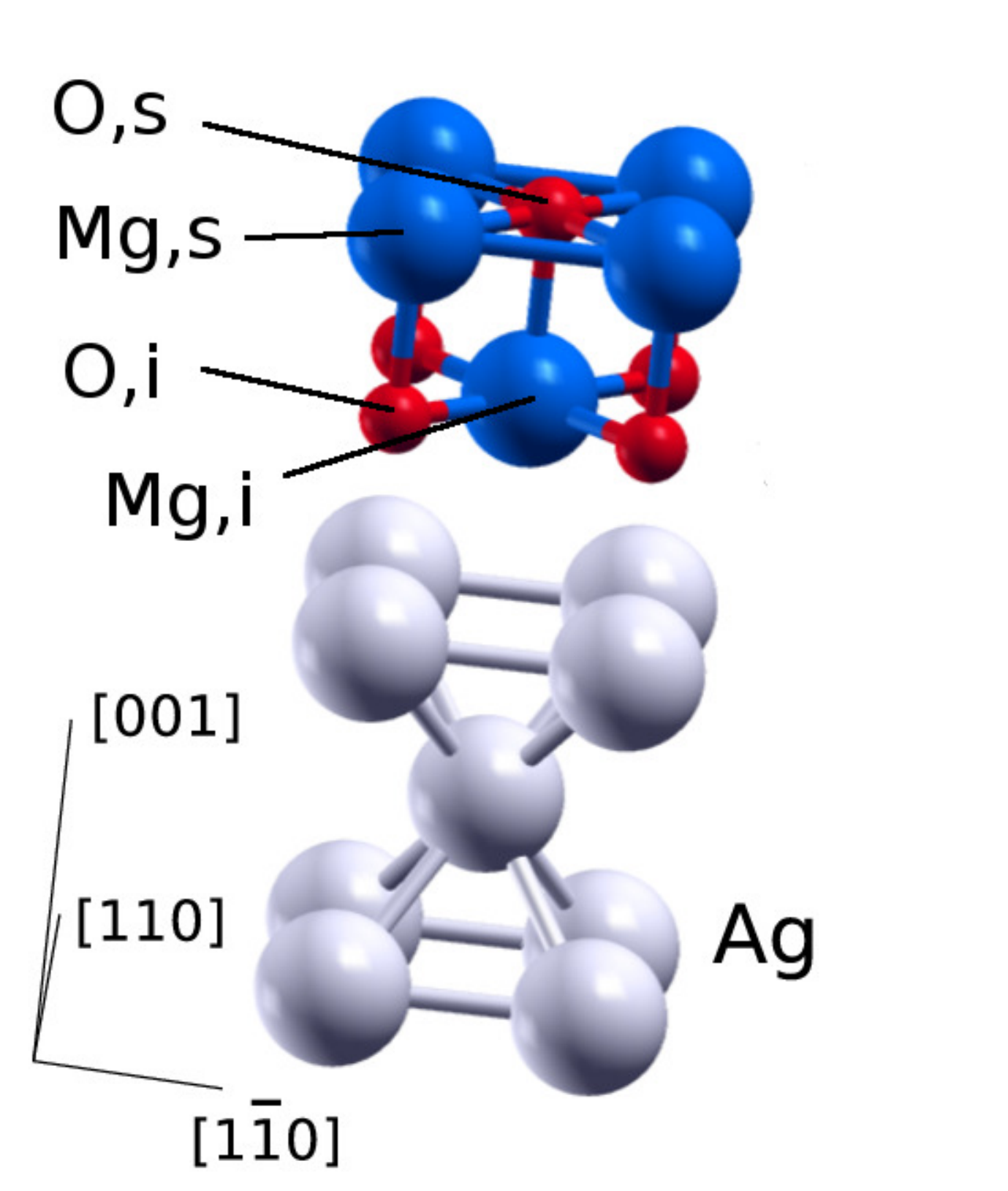}
\caption{(Color online) The unit cell describing 2 ML of MgO
on top of 3 ML of Ag (001): `i' means `interface' and `s' stands for `surface'.  The in-plane   [110] and [1$\bar1$0] directions are labeled as is the out-of-plane [001] direction.}
\label{fig:slab}
\end{figure}

Next, we studied an epitaxial ultrathin MgO film (2 ML) on  the silver substrate (3 ML).
The unit cell for this calculation is shown in Fig.~\ref{fig:slab} as are the final relaxed configuration: the O atoms are directly above Ag atoms, a configuration that 
is energetically more favorable than the alternative alignments (such as placing Mg on top of Ag).
The structural parameters of the two bottom silver layers were fixed
to the bulk values, while the coordinates of the MgO atoms and the interfacial Ag were fully relaxed (within the epitaxial constraint).  Numerical coordinate data on the relaxed structure is displayed in Table~\ref{tab:slab_coords}.
Our tabulated results are in agreement with the theoretical study of 
Lopez {\it et al.},\cite{lopez_04}
aside from the substrate-overlayer distance $d\equiv z_{\mathrm{O,i}}-z_{\mathrm{Ag}}$ which in our case is about $0.2$~\AA\ larger than theirs.  However, Giordano {\it et al.}\cite{giordano_06}
found a substrate-overlayer distance $d=2.73$ \AA\ in this system which is very similar to our result.  Unfortunately, as we show below, the detailed STM image depends strongly on the precise value of $d$ which serves as a caution to blind comparison to experiment without careful benchmarking in this specific system.  We describe the reason for the $d$ dependence in Section~\ref{sec:model}.

\begin{table}
  \caption{\label{tab:slab_coords}
 Calculated relaxed geometry for 2 ML of MgO on  Ag (001). 
 All coordinates are in \AA, `i' stands for `interface',
 and `s' stands for `surface'.  `Ag' refers to the interfacial silver atom (see Fig.~\ref{fig:slab}), and $z_{\mathrm{Ag}}-z_{\mathrm{Ag,bulk}}$ describes the vertical relaxation of the interfacial Ag compared to its bulk unreconstructed position on the surface.
 }
\begin{ruledtabular}
\begin{tabular}{lll}
  & This work & Previous theory \\
  $z_{\mathrm{Ag}}-z_{\mathrm{Ag,bulk}}$ & $-0.05$ &
                     $-0.06$\footnotemark[1] \\  
      $z_{\mathrm{O,i}}-z_{\mathrm{Ag}}$ & $\phm{-}2.68$ &
                     $\phm{-}2.47$\footnotemark[1], $2.73$\footnotemark[2]\\
    $z_{\mathrm{Mg,i}}-z_{\mathrm{Ag}}$  & $\phm{-}2.67$ &
                     $\phm{-}2.45$\footnotemark[1] \\  
    $z_{\mathrm{O,s}}-z_{\mathrm{Mg,i}}$ & $\phm{-}2.21$ &
                     $\phm{-}2.20$\footnotemark[1] \\ 
    $z_{\mathrm{Mg,s}}-z_{\mathrm{O,i}}$ & $\phm{-}2.17$ &
                     $\phm{-}2.14$\footnotemark[1]
\end{tabular}
\footnotetext[1]{Ref.~\onlinecite{lopez_04}.}
\footnotetext[2]{Ref.~\onlinecite{giordano_06}.}
\end{ruledtabular}
\end{table}
\begin{figure}
\centering\includegraphics{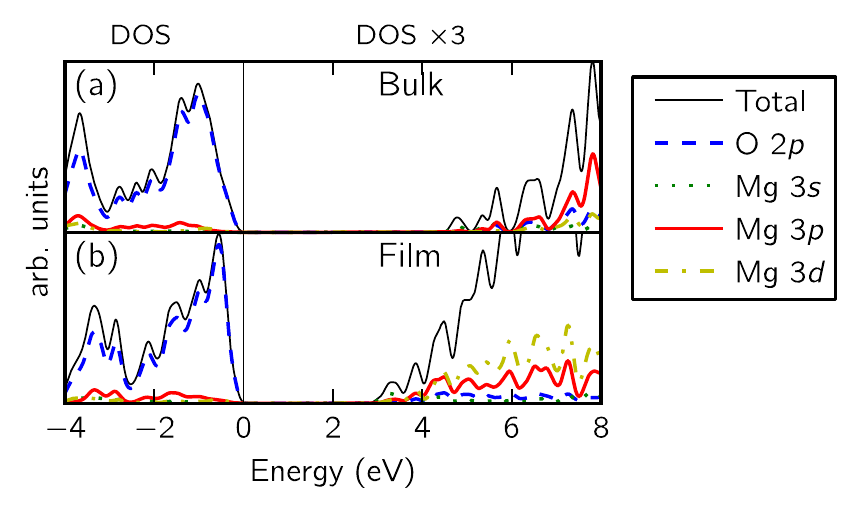}
\caption{
Total (DOS) and projected density of states (PDOS) of (a) bulk MgO
and (b) a 2 ML free standing MgO film in vacuum. The energy
is referenced to the top of the valence band indicated by the vertical black line
at zero energy. To help visualize the conduction band states, DOS and PDOS are multiplied by a factor of 3 for energy above zero.
}
\label{fig:pdos_mgo}
\end{figure}
\begin{figure}
\centering\includegraphics{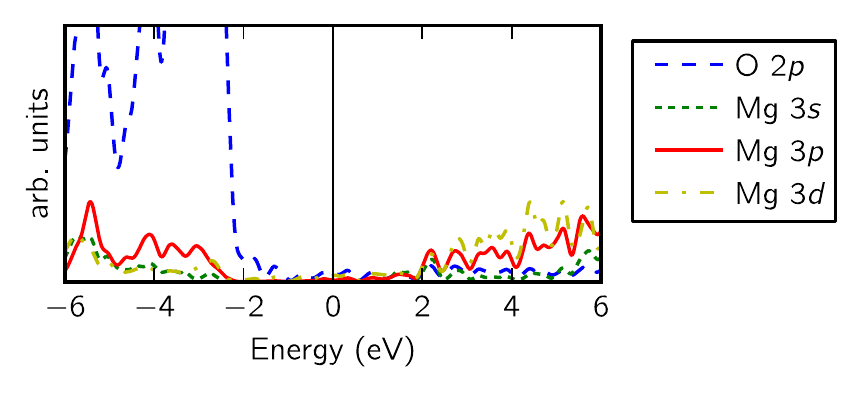}
\caption{
PDOS on MgO atomic orbitals of the 2 ML MgO/Ag interfacial system.  Energies are referenced to the Fermi level which is set to zero energy.
}
\label{fig:pdos_ag_mgo}
\end{figure}
To examine the electronic structure of this interfacial system and the resulting STM image, it is  fruitful to consider in parallel the isolated subsystems: a 2 ML MgO thin film in vacuum and an isolated Ag substrate (ML) in vacuum.  For these systems, the atomic coordinates are fixed at those derived from the relaxation of the original interfacial system.  Figures~\ref{fig:pdos_mgo}(a) and (b) compare the DOS and PDOS for bulk MgO and the MgO thin film.
We note that the band gap of the MgO film (3.0 eV) is smaller than the bulk (4.4 eV), a result consistent with previous calculations as well as experimentally observed band gap
reduction in a thin film compared to bulk.\cite{schintke_01}  By comparison, Fig.~\ref{fig:pdos_ag_mgo} shows the PDOS of the MgO subsystem in the actual MgO/Ag interfacial system: the PDOS in the MgO subsystem is now non-zero in the band gap region of the free standing MgO film, and specifically non-zero PDOS develops at the Fermi level due to the coupling to the Ag substrate.  This result already hints at the fact that states in the MgO may play a role in the formation of the STM image.

\begin{figure}
  \centering\includegraphics[width=250pt]{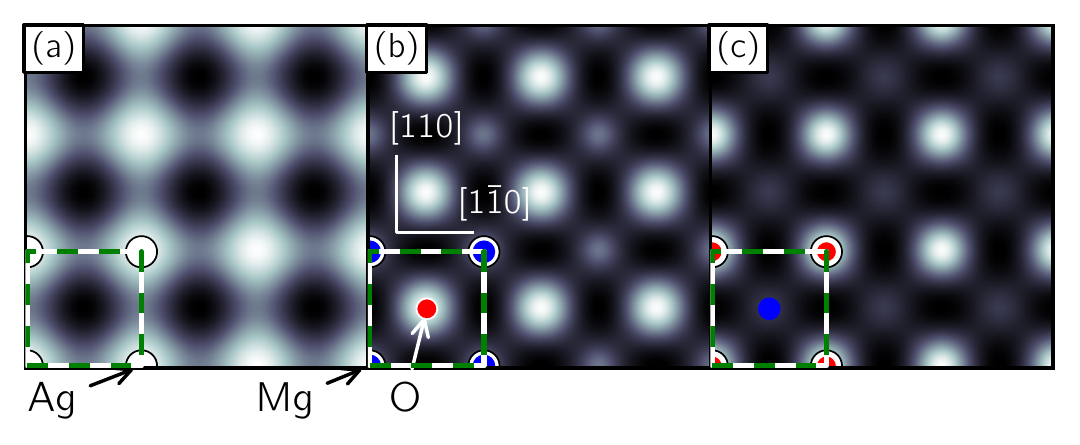}
\caption{
Constant-height near-zero-bias STM simulations of surfaces of (a) Ag (001),
(b) 2 ML MgO/Ag(001), and (c) 3 ML MgO/Ag(001). 
The images are computed at $1$~\AA\ above the surfaces.
The blue (red) circles denote the positions of Mg (O) atoms at the surface
monolayer. The white circles denote positions of Ag atoms
of the interfacial layer.
The dashed squares outline the unit cell used in these calculations.
}
\label{fig:mgo_ag_stm}
\end{figure}

We now proceed to the calculated STM images for this system.  The STM image (LDOS at \EF) for the bare Ag (001) surface is shown in 
Fig.~\ref{fig:mgo_ag_stm}(a): not surprisingly, the bright spots 
correspond to the locations of the silver atoms on the surface layer and follow the square pattern of the surface lattice.  The STM image for the relaxed MgO/Ag system is shown in Fig.~\ref{fig:mgo_ag_stm}(b): the brightest spots in the STM image are at the location of the surface O atoms with weaker features at the Mg locations.  In Fig.~\ref{fig:mgo_ag_stm}(c), we show the result for 3 ML of MgO on Ag (001): the bright features continue to track the positions of the surface O atoms, and the overall image simply looks like a translated version of the 2 ML result of Fig.~\ref{fig:mgo_ag_stm}(b).
We also note that when we calculate the STM image of the bare Ag (001) surface at the same height above the top Ag atoms as in the 2 ML MgO/Ag calculation, we obtain essentially a null result: the maximum density is smaller by a factor of $10^2$ than the maximum density in Fig.~\ref{fig:mgo_ag_stm}(b) and there is almost no contrast.
 In addition to the in-plane shift of the image, the overall intensity of the STM image for 3 ML is smaller than for 2 ML due to the exponential decay of the states at the Fermi level through the MgO film, a fact we have verified by studying a range of thicknesses from 2 to 8 ML of MgO.  Since the Ag is obviously crucial to having states at the Fermi level, and yet the STM image tracks the position of the surface atoms of the insulator, a satisfactory explanation for low-bias observations is lacking and motivates further analysis and computations below.

\begin{figure}
  \centering\includegraphics[width=250pt]{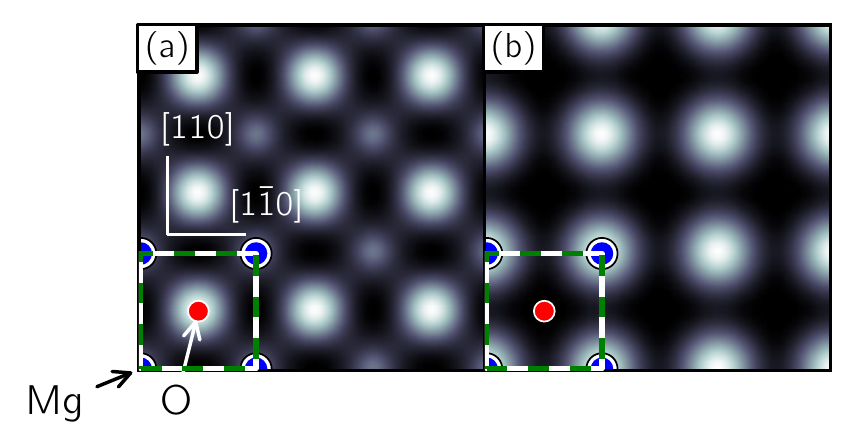}
\caption{
Dependence of STM images on the overlayer-substrate distance $d$ for 2 ML MgO/Ag (001).  (a) is based on our theoretically relaxed structure with $d=2.68$~\AA\ and is identical to Fig.~\ref{fig:mgo_ag_stm}(b).  (b) uses $d$ fixed to 2.47~\AA. STM images are computed at $1$~\AA\ above the surfaces. The blue (red) circles denote the positions of Mg (O) atoms at the surface
monolayer. The dashed squares outline the unit cell used in the calculations.
}
\label{fig:mgo_ag_stm_fd}
\end{figure}

These STM images do not agree with those given by Lopez~{\it et al.}\cite{lopez_04} who observed bright spots at locations on top of the surface Mg atoms which vertically overlap the interfacial Ag atoms in the 2 ML MgO case.\cite{explan_01}  However, there is no great cause for concern due to the strong overlayer-substrate separation $d$ dependence of the STM results.  In 
Fig.~\ref{fig:mgo_ag_stm_fd}(c) we show the STM image computed by using $d$ fixed at the value taken from Ref.~\onlinecite{lopez_04}.  The STM image in this case now resembles their result with bright spots at the surface Mg locations.  

Before ending this section, we briefly describe some high-bias-voltage results for the Mg/Ag system.  We simulated the STM image at 3~V bias for MgO thickness of 1, 2, 3, 4, 8, and 16 ML.  The STM features are always located at the positions of the Mg atoms on the topmost layer, and staring at 2 ML, the STM image intensity does not show any dependence on the MgO thickness. Combined with the PDOS results in Fig.~\ref{fig:pdos_mgo}, we conclude that, as noted previously,\cite{lopez_04} the high-bias-voltage STM probes the Mg-derived conduction bands of the MgO film itself.  We note that even though Tersoff-Hamann approximation is not very well justified for high-bias voltages, it yields a physically reasonable
and qualitatively correct result for this system.

\subsection{Overlayer shift}
\label{sec:shift}

\begin{figure}
  \centering\includegraphics[width=250pt]{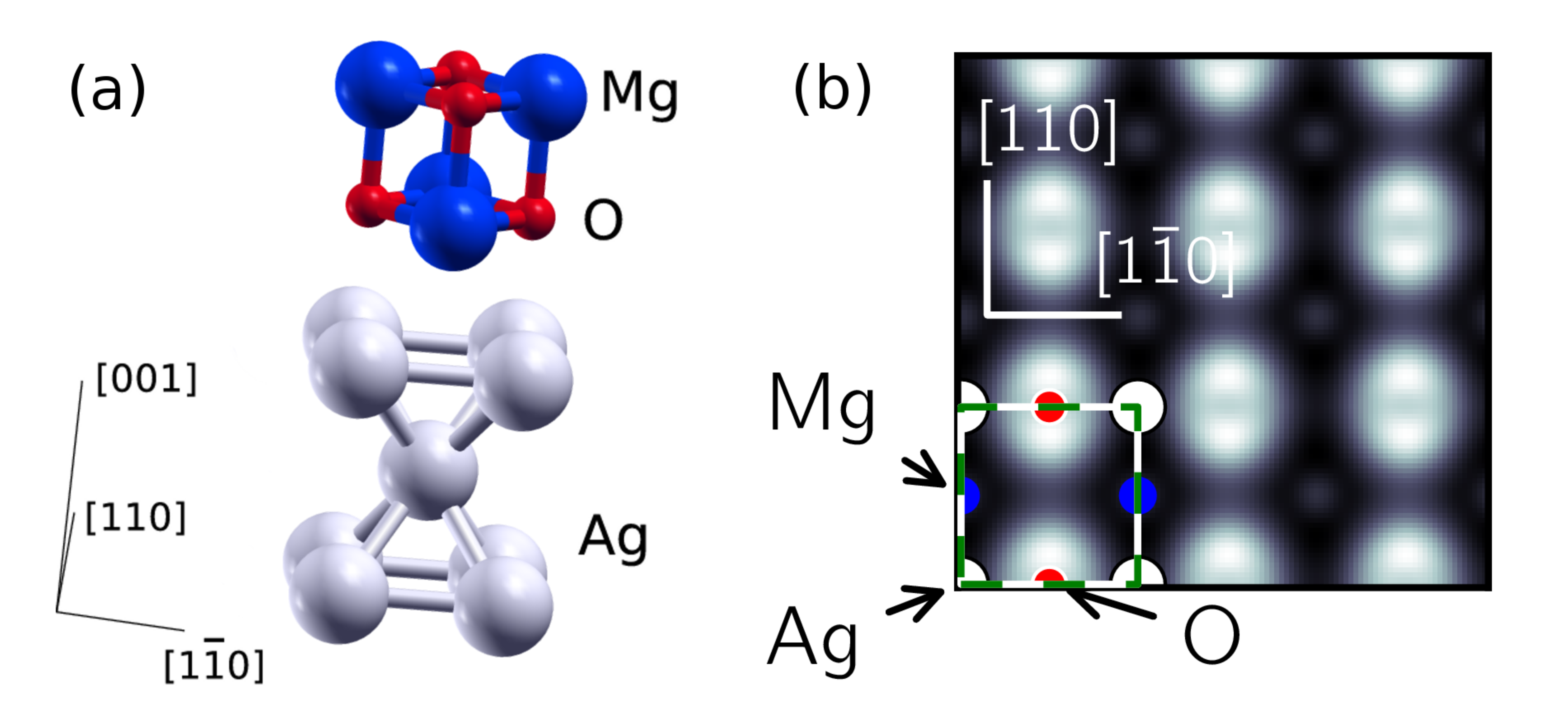}
\caption{
  (Color online) (a) Side view of the 2ML MgO/Ag(001) system with
  the MgO overlayer shifted horizontally in the $[110]$ direction
  by half of the unit cell from its relaxed position.
  (b) Corresponding constant-height near-zero-bias STM
  image simulated at $1$~\AA\ above the surface.
 The blue (red) circles denote the positions of Mg (O) atoms at the surface
 monolayer. The white circles denote positions of Ag atoms
 at the interfacial layer.
  The dashed squares outline the unit cell used in the simulation.
}
\label{fig:mgo_shift}
\end{figure}

In a next step, we consider ``sliding'' the MgO layer with respect to the Ag substrate.  In particular, starting from the original unit cell shown in Fig.~\ref{fig:slab}, we shift the MgO overlayer by half of the unit cell in the $[110]$ direction.  In the resulting structure shown in Fig.~\ref{fig:mgo_shift}(a), the in-plane coordinates of the MgO atoms do not coincide with the Ag ones.  The resulting low-bias STM image shown in Fig.~\ref{fig:mgo_shift}(b) clearly indicates that the bright spots in the STM image move together with the overlayer atoms (O atoms in particular) with no visible features at the positions of the Ag atoms.

\subsection{Incommensurate overlayer}
\label{sec:incommensurate}

\begin{figure*}
  \centering\includegraphics[width=400pt]{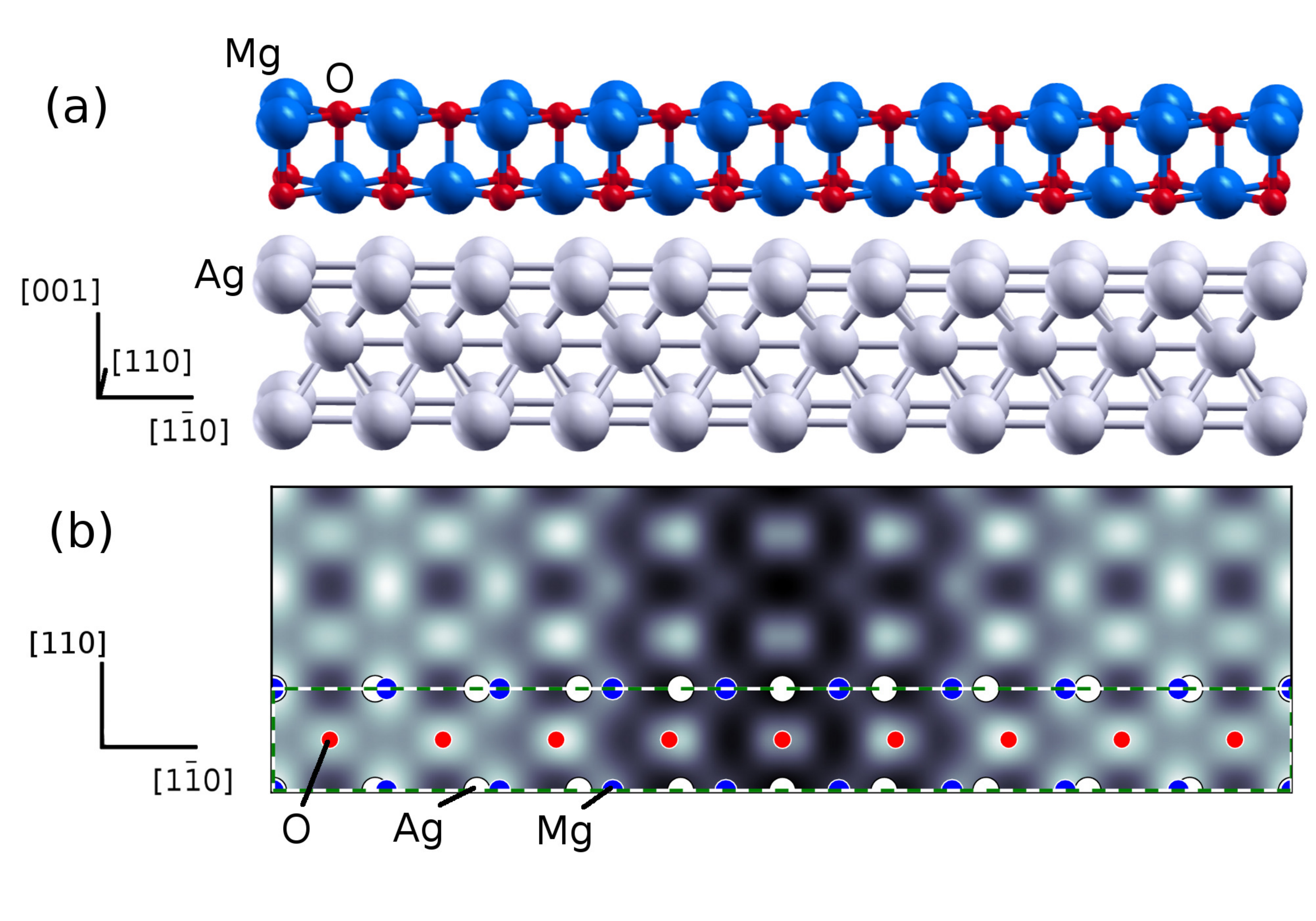}
\caption{
 (a) Side view of the 2ML MgO/Ag(001) system with
  the MgO overlayer stretched horizontally.
  (b) Corresponding constant-height 
  near-zero-bias STM image. The dashed rectangle outlines the unit cell
  used in the simulation.
}
\label{fig:wave}
\end{figure*}

To further elucidate the origin of the STM features in MgO/Ag system at low-voltage bias,
we construct a number of simulation cells where the relative in-plane registry of the MgO and Ag atoms varies across the unit cell which allows us to directly examine the effect of the relative alignment of the two materials on the STM image in a single calculation.  

We consider a supercell which is elongated to contain 10 primitive lattice vectors of the bulk Ag substrate in the horizontal direction.  We then uniformly stretch a 2 ML thick MgO film which is 9 unit cells long in the horizontal direction to lie on top of the 10 unit cells of Ag (tensile strain).  No structural relaxations are performed and all vertical coordinates are fixed to those quoted in Table~\ref{tab:slab_coords}.\cite{explan_02}  The resulting structure is shown in Fig.~\ref{fig:wave}(a) along with the resulting STM image in Fig.~\ref{fig:wave}(b).  If the Ag atoms dominated the STM current at low bias, we would expect to see bright features with the periodicity of the Ag substrate which is clearly not the case.  Instead, the maxima of the LDOS at \EF track the periodicity of the MgO overlayer.  In addition,
there is an overall Moir\'e-like modulation of the brightness that indicates the presence of interference between various contributions to the final STM image.
The Moir\'e pattern is brightest in places where the MgO overlayer
is in registry with the substrate --- when the interfacial O atoms lie close to the the interfacial Ag atoms.
Similar Moir\'e patterns have been observed experimentally in STM images of mismatched oxide layers on metal surfaces.  Because it is not possible to unambiguously distinguish topographic from electronic contrast in STM images, the contrast in these experiments may be due to a rumpling of the oxide layer as it moves in and out of registry with the substrate or due to a change in coupling of the oxide layer to the metal as the registry varies.\cite{galloway_93,li_14}

\begin{figure*}
  \centering\includegraphics[width=400pt]{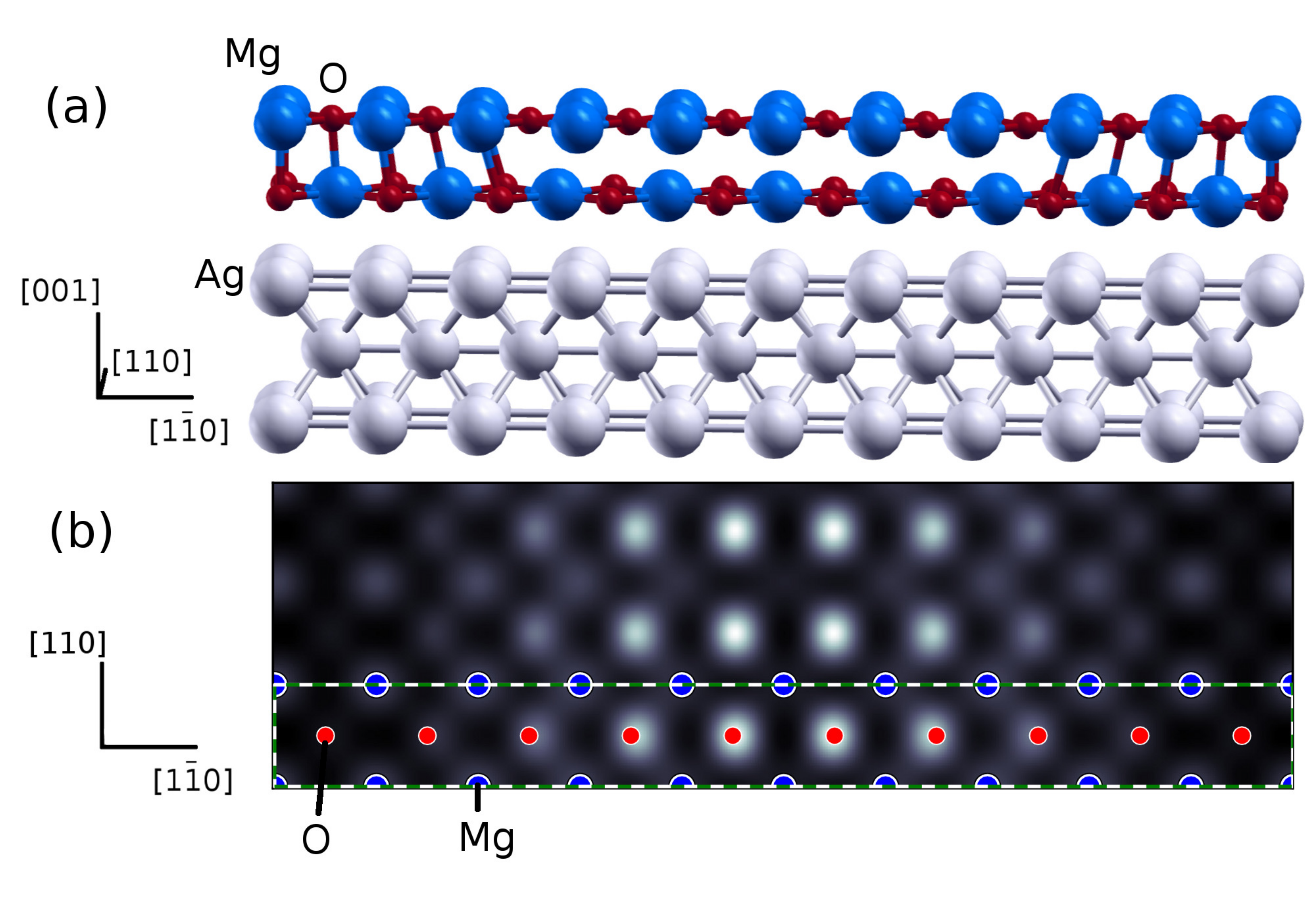}
\caption{
(a) Side view of the 2ML MgO/Ag(001) system with
  only the interfacial MgO layer stretched horizontally.
  (b) Corresponding constant-height 
  near-zero-bias STM image. The dashed rectangle outlines the unit cell
  used in the simulation.
}
\label{fig:mgo_semiwave}
\end{figure*}

In an attempt to separate the role of each of the two MgO monolayers, we performed a new calculation where we stretched only the atomic plane in contact with the Ag substrate while keeping the MgO layer on the surface in registry with the Ag substrate.  The resulting structure and STM image are shown in Figs.~\ref{fig:mgo_semiwave}(a) and (b).  In this case, the bright STM features have the periodicity of the top MgO layer, but the location of the brightest features change between Mg and O surface atoms as a function of the horizontal coordinate. 

Broadly speaking, the above two calculations provide the following picture.  The STM image is formed from some non-trivial hybridization of the Ag with the Mg and O atoms in the interfacial region.  The hybridization is significant even when the Mg and O atoms are out of registry with the Ag atoms, but the brightest STM images occur when the atoms are in registry as one would expect in such a case a maximized overlap of atomic orbitals between Ag and MgO.  We now examine each of two results in more detail.

We consider more closely the structure shown in Fig.~\ref{fig:wave}(a).
In the left corner of the unit cell, we have an interfacial O atom placed
directly above the interfacial Ag atom, and the surface Mg atom is above the interfacial O atom.  This represents a case of maximum overlap of atomic orbitals which leads to the brightest features in the STM image of Fig.~\ref{fig:wave}(b).  On the other hand, in the center of the unit cell there are no Mg or O atoms directly on top of the Ag atoms which diminished overall STM brightness.  In addition, the surface O atoms appear brighter in the central location which indirectly highlights their relative importance in forming the STM image.  However, the intensity pattern is formed via a complex interaction in the interfacial region:  the brightest features in Fig.~\ref{fig:mgo_semiwave}(b) occur in the center of the unit cell where the surface O atoms
are located directly above the Ag atoms in the {\it second} layer from the silver surface and there are no other atoms in between. This foreshadows the fact that a straightforward analysis of the Ag to MgO coupling that leads to the STM features is a complex undertaking.

\begin{figure}
  \centering\includegraphics[width=250pt]{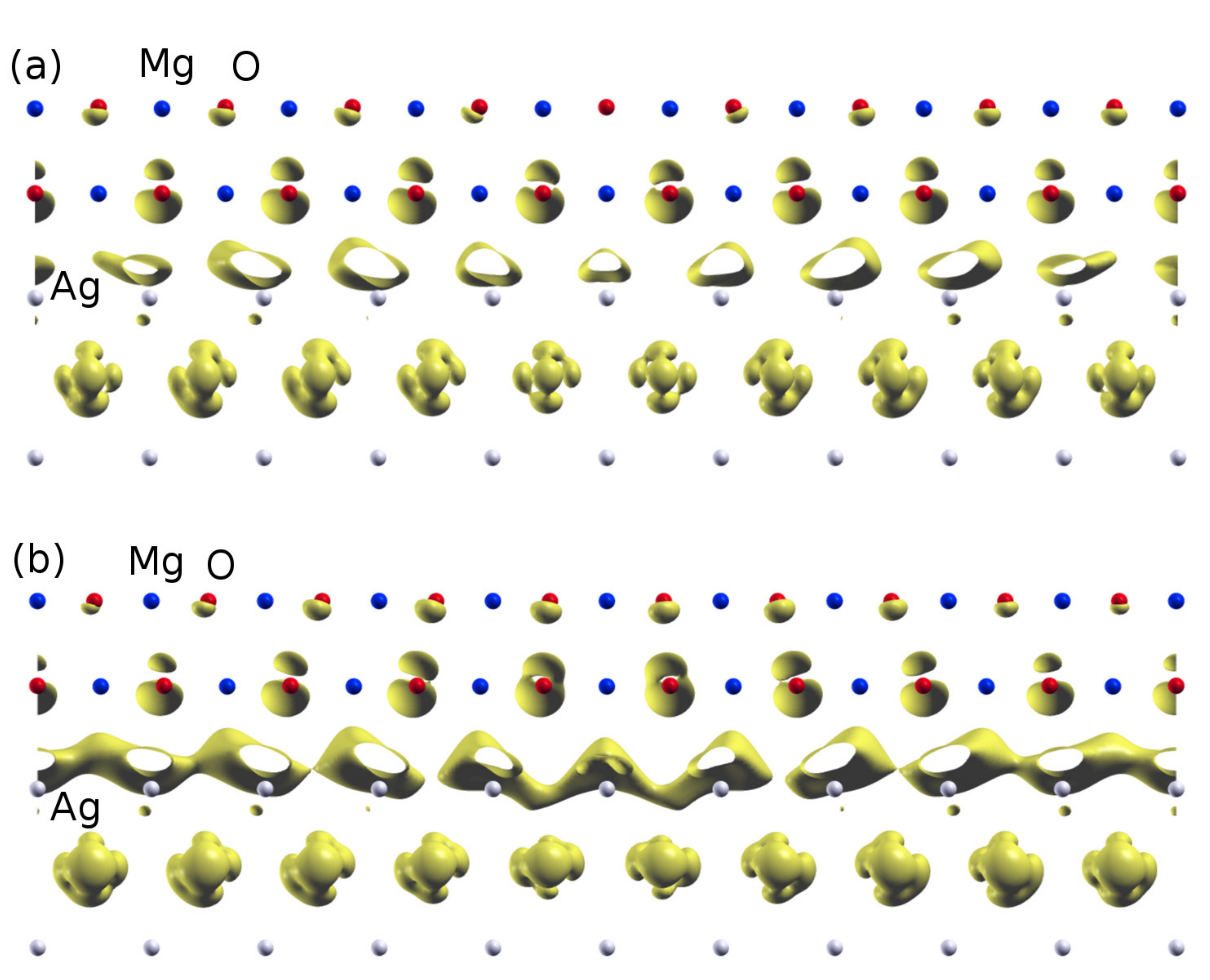}
\caption{
  Local density of states at the Fermi level for the structures
  shown in Fig.~\ref{fig:wave}(a) and Fig.~\ref{fig:mgo_semiwave}(a).
  The isosurface is shown at $\sim10\%$ of the maximum value of the function.
}
\label{fig:migs_3d}
\end{figure}
To illustrate the importance of the oxygen sublattice in the formation of the LDOS at \EF and hence the STM image, we display three dimensional isosurface plots of the LDOS at \EF in Fig.~\ref{fig:migs_3d} for the two structures discussed in this section. Inside the MgO, we see primarily 
a decaying evanescent behavior localized on the oxygen sublattice dominated by O $2p_z$ orbitals.  What this figure makes completely clear is that that the propagation of the evanescent states throughout the MgO is through the tight-binding orbitals of the MgO lattice and not as free-electron like states originating at the Ag surface.  Based on this finding, we will be performing a more thorough analysis of the MgO/Ag system using a tight-binding representation.

\section{Tight-binding model}
\label{sec:model}

While direct first-principles computation of the STM image (LDOS at \EF in our case) is the final theoretical output that may be compared with experiment, the mechanism that leads to the formation of the STM image on the surface is difficult to tease out from such direct computations of the final result.  For example, it is difficult to simply and selectively ``turn off'' the contribution of certain atoms to the STM image without changing their mutual couplings or couplings to other atoms (or vice versa).  A tight-binding description, on the other hand, allows one to answer a number of questions of this variety by providing a number of additional analysis opportunities.  

We construct a tight-binding model of the MgO/Ag system as follows.  Starting with the relaxed 2 MgO/Ag system, we perform a self-consistent field calculation while sampling the Brillouin zone on a fine $24\times24$ mesh of $\k$ points (the high density of $\k$ points is necessary for a smooth Fourier interpolation of the band structure in the L\"owdin basis).  Armed with the Hamiltonian and overlap matrices at each $\k$ point, $H^\k_{\alpha\beta}$ and $S^\k_{\alpha\beta}$, where Greek letters label atomic orbitals in one unit cell, we Fourier transform them to find a real-space tight-binding description.  For example, the real-space Hamiltonian $h^\R_{\alpha\beta}$ is computed via 
\begin{equation*}
\label{eq:HR}
h^\R_{\alpha\beta}=\frac1{N_k}\sum_{\k}H^\k_{\alpha\beta}e^{-i\k\cdot\R},
\end{equation*}
where $N_k$ is the total number of $\k$ points ($24^2=576$ in this case) and $\R$ is a lattice vector. (An analogous formula connects $S^\k$ to $s^{\R}$.) The matrix element 
$h^\R_{\alpha\beta}$ is that between orbital $\alpha$ in the ``home'' unit cell at the origin to orbital $\beta$ in the unit centered at position $\R$.  Diagonal elements $h^0_{\alpha\alpha}$ are the on-site energies of the orbitals.  While it is generally preferable to use a tight-binding basis formed from maximally localized Wannier functions\cite{marzari_97} which automatically guarantees orthonormality, compactness, and exact representation of the Hilbert subspace of interest, the procedure to generate such Wannier functions is extremely difficult (and was abandoned) for this material system due to the very large spread of the Mg atomic orbitals described below.  Finally, we use Fourier interpolation to generate $H^\k$ at an arbitrary $\k$ point not in the original grid:
\begin{equation*}
H^\k_{\alpha\beta}=\sum_{\R}h^\R_{\alpha\beta}e^{i\k\cdot\R}
\end{equation*}
(and analogously for getting $S^\k$ from $s^\R$). Solving the generalized eigenvalue problem to find the energy bands $E_{n\k}$,
\begin{equation*}
\sum_{\beta} H^\k_{\alpha\beta} C^\k_{\beta n} = E_{n\k}\sum_\beta S^\k_{\alpha\beta}C^\k_{\beta n}\,,
\end{equation*}
we generate the electronic band structure within our tight-binding model.  

\begin{figure}
  \centering\includegraphics[width=250pt]{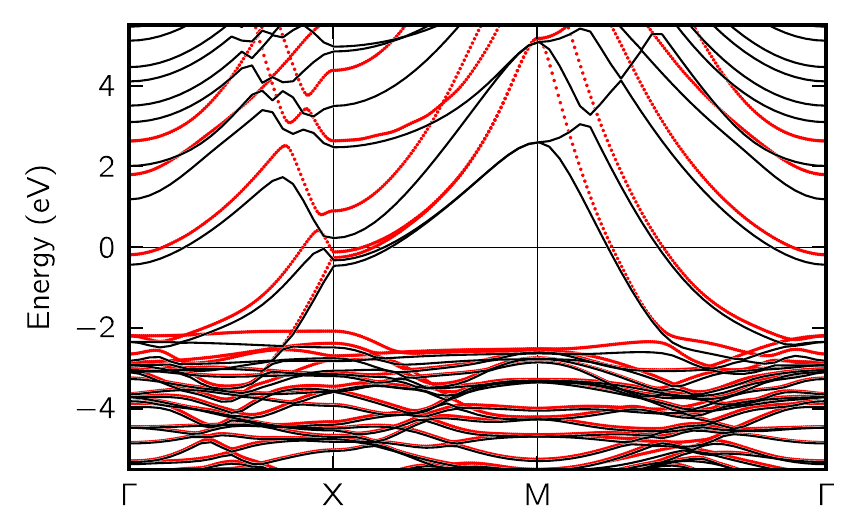}
\caption{
Band structure versus in-plane wave vector $\k$ for 6 ML of MgO on  Ag (001).
(Black) solid lines show the plane wave basis band structure
  and (red) dotted lines show the tight-binding band structure. The Fermi level is placed
  at 0 eV.
}
\label{fig:tb_bands}
\end{figure}

The resulting band structure for 2 ML MgO/Ag is shown in Fig.~\ref{fig:tb_bands}
together with the original plane-wave basis band structure.
The tight-binding basis provides a good quantitative description of the valence bands as well as the bands that are at or near the Fermi energy.  The degradation in quality above the Fermi level is normal and expected of a method based on localized orbitals once the eigenstates contain significant delocalized plane-wave character, which is the case in the conduction band at high energies. As we are focused on the states at or near the Fermi level, which form the low-bias STM image, the present tight-binding approach is deemed sufficient for continued analysis.

We use the tight-binding model to compute the STM image by computing the LDOS at \EF at point \rb 
\begin{equation*}
L(\rb,E_{\mathrm{F}})=\sum_{n,\k}w(E_{\mathrm{F}}-E_{nk})\left\vert\psi_{nk}(\rb)\right\vert^2,
\end{equation*}
where the window function $w(E_{\mathrm{F}}-E_{nk})$
selects the states $\vert\psi_{nk}\rangle$ with energies $E_{nk}$
in the vicinity of the Fermi level $E_{\mathrm{F}}$ (for the purpose of computing the STM images shown below, 
we used the window function selecting the states at $E_{\mathrm{F}}\pm0.1$~eV).
 The Bloch state $\psi_{nk}(\rb)$ is expanded in the basis of atomic orbitals $\phi_\alpha(\rb)$ via
\beq
\psi_{n\k}(\rb) = \frac{1}{\sqrt{N_k}}\sum_{\R,\alpha} C^\k_{\alpha n} e^{i\k\cdot\R}\phi_\alpha(\rb-\R)\,.
\eeq
Written explicitly in terms of the expansion coefficients $C^\k_{\alpha n}$ and atomic orbitals, 
\beq
\label{eq:tb_ldos}
L(\rb,E_{\mathrm{F}})=\frac1{N_k}\sum_{n,\k}w(E_{\mathrm{F}}-E_{nk})\left\vert
\sum_{\alpha,\R }C^\k_{\alpha n}\phi_{\alpha}(\rb-\R)\right\vert^2.
\eeq
By using the atomic orbitals $\phi_\alpha(\rb)$ used in the generation of the pseudopotentials together with the energies $E_{n\k}$ and coefficients $C^\k_{\alpha n}$ from the tight-binding model, we can compute the LDOS at \EF for any point $\rb$ above the surface and thus compute an STM image based on tight-binding theory.

\begin{figure}
  \centering\includegraphics[width=250pt]{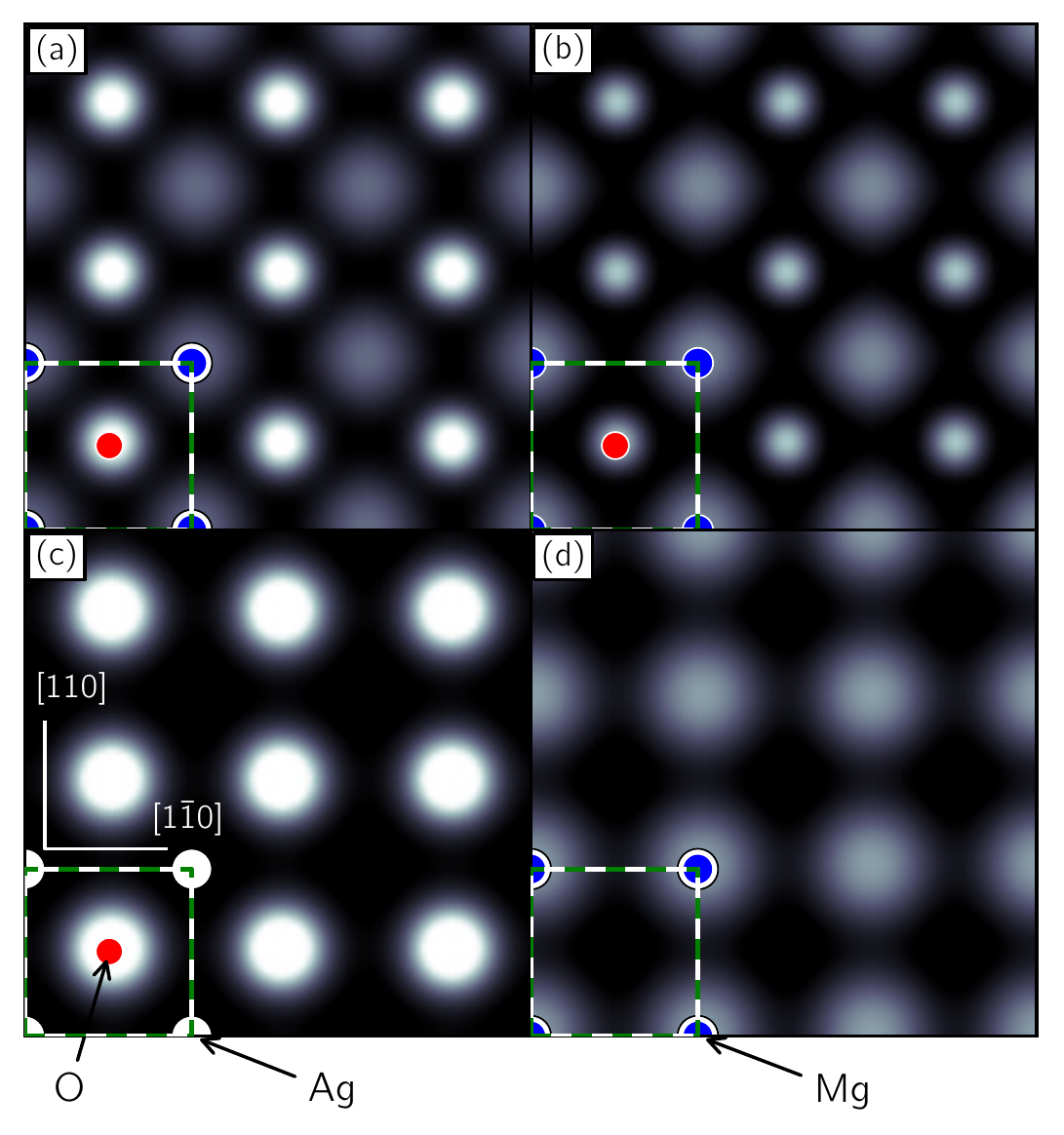}
\caption{
 Constant-height LDOS simulations of the 2 ML MgO/Ag(001) system
 based on the atomic-orbital tight-binding model and selective omission of orbitals in the computation of the LDOS at \EF in Eq.~(\ref{eq:tb_ldos}):
 (a) all orbitals included, 
 (b) orbitals of all Ag atoms omitted,
 (c) orbitals of the surface Mg omitted, and
 (d) orbitals of the surface O omitted.
 All images are computed at $1$~\AA\ above the surface.
 The blue (red) circles denote the positions of Mg (O) atoms on the surface layer.
  The white circles denote positions of Ag atoms
 at the interfacial layer.
 The dashed squares outline the unit cells used in the calculations.
}
\label{fig:tb_stm}
\end{figure}

The LDOS computed with Eq.~(\ref{eq:tb_ldos}) will always be an approximation to the more exact answer given by the computation of $\psi_{n\k}(\rb)$ using plane waves, and so the main question is to what extent we can use the LDOS from the tight-binding model to understand the electronic behavior of the system.  Figure~\ref{fig:tb_stm}(a) shows the LDOS at \EF for 2 ML MgO/Ag system based on Eq.~(\ref{eq:tb_ldos}). One can see that qualitatively
it looks very similar to the image shown in Fig.~{\ref{fig:mgo_ag_stm}}(b), and this means that
the atomic orbital basis is complete enough in the regions of interest above the surface to generate STM images that match the plane-wave results in terms of key features and brightness ratios.  This encourages us to use this tool for further analysis.

One advantage of having a tight-binding model is the ability to separate contributions from different orbitals to the STM image.  For example, Figure~\ref{fig:tb_stm}(b) shows the LDOS at \EF from Eq.~(\ref{eq:tb_ldos}) where all coefficients $C^\k_{\alpha n}$ appearing in that equation that correspond to atomic orbitals on all Ag atoms are set to zero.  The fact that the image is hardly changed from the case that includes all orbitals [Fig.~\ref{fig:tb_stm}(a)] unambiguously proves that the STM contrast is formed by electronic states inhabiting the orbitals of the MgO.  Next, we can zero contributions from selected surface atoms and observe changes in the STM image.  Fig.~\ref{fig:tb_stm}(c) shows the LDOS at \EF when surface Mg orbitals are omitted, and Fig.~\ref{fig:tb_stm}(d) shows the resulting image when surface O orbitals are omitted.  Clearly, the strong changes in the images shows that the surface atoms dominate the STM image formation.  Furthermore, we note that omitting the Mg orbitals increases the overall brightness --- this points out the fact that the final STM image is formed by a superposition of mainly destructively interfering contributions from the Mg and O sublattice above the surface.

\begin{figure}
  \centering\includegraphics[width=250pt]{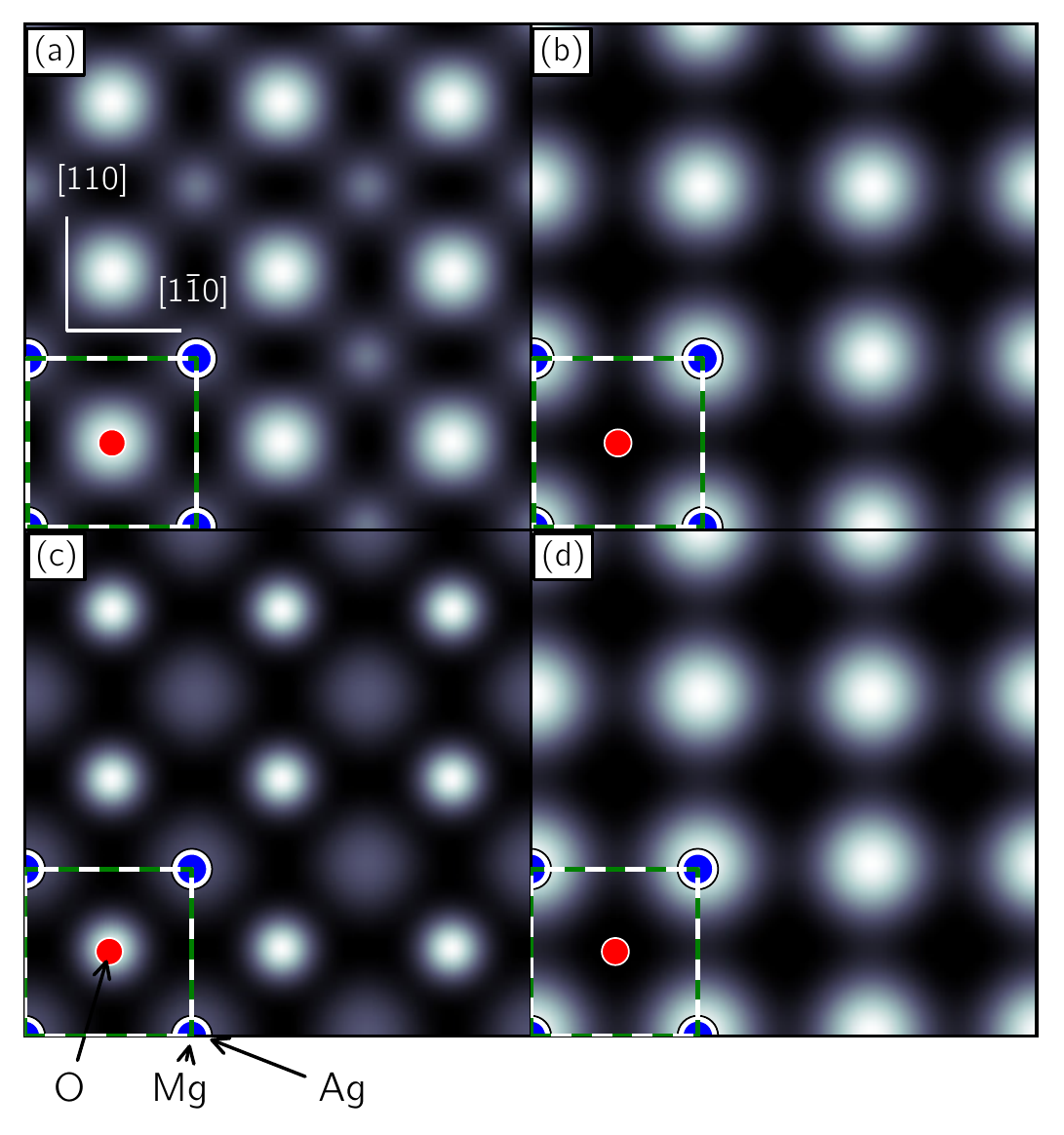}
\caption{
 Constant-height LDOS simulations based on the plane-wave theory [panels (a) and (b)] and the tight-binding model 
 [panels (c) and (d)] for the 2 ML MgO/Ag(001) system.
 Panels (a) and (c) correspond to a relaxed geometry of the interface with MgO/Ag separation distance $d=2.68$~\AA, while
 panels (b) and (d) correspond to fixed $d=2.47$~\AA.
 The images are computed at $1$~\AA\ above the surface.
 The blue (red) circles denote the positions of Mg (O) atoms at the surface
 monolayer. The white circles denote positions of Ag atoms
 at the interfacial layer.
 The dashed squares outline the unit cells used in the calculations.
}
\label{fig:tb_vs_dft}
\end{figure}

Applying the above methodology to the case where the MgO/Ag separation $d$ is fixed at 2.47~\AA, we obtain the image shown in Fig.~\ref{fig:tb_vs_dft}(d) which, again, is in agreement with the plane-wave result of Fig.~\ref{fig:mgo_ag_stm_fd}(b) [reproduced in Fig.~\ref{fig:tb_vs_dft}(c) for convenience along with the plane-wave STM simulations in panels (a) and (b)]: the Mg sites become much brighter.
The tight-binding method is thus able to reproduce the $d$ dependence of the STM contrast and can be used to understand why this happens, as detailed below.

\begin{figure}
  \centering\includegraphics[width=240pt]{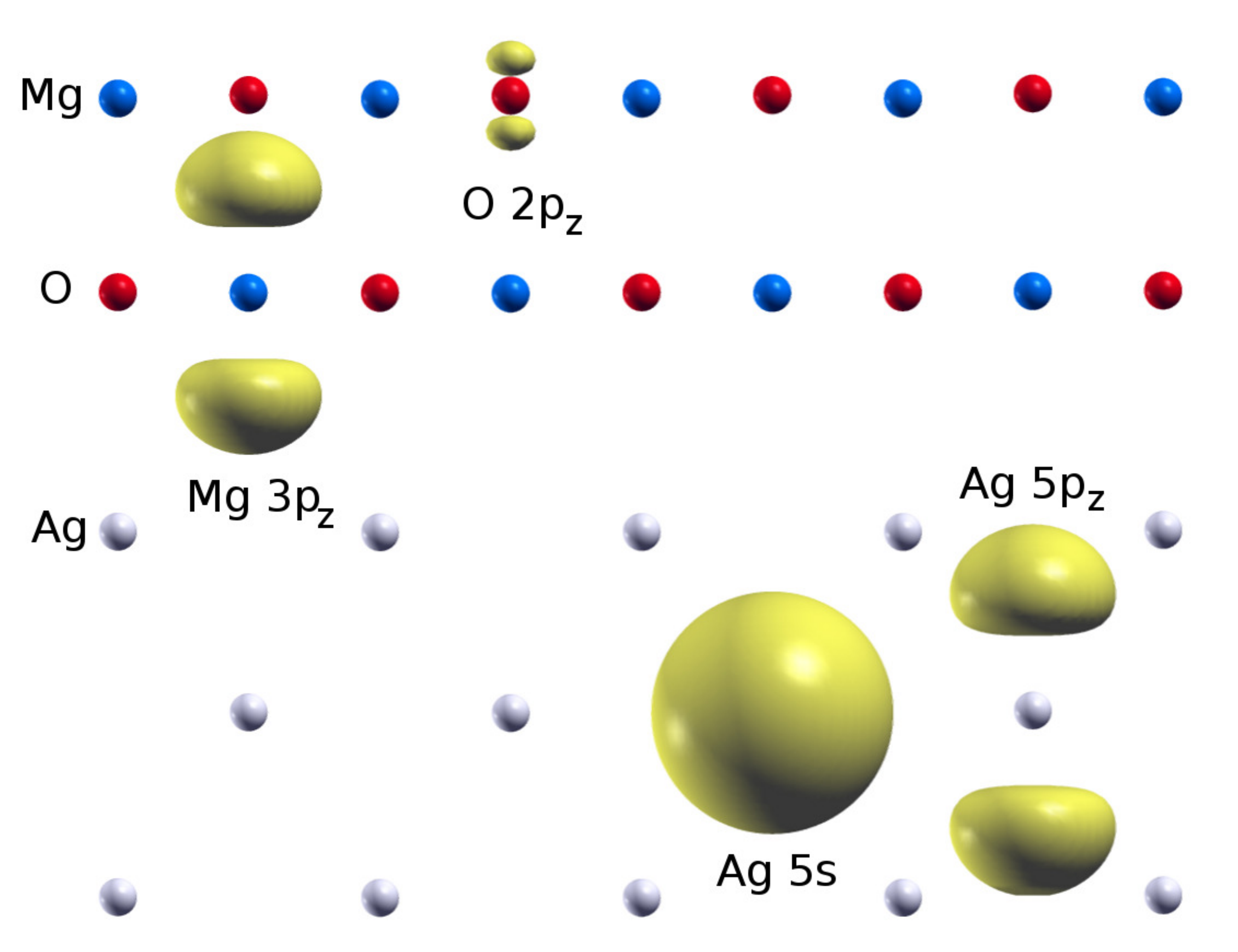}
\caption{
Isosurfaces of the square of the wave functions for the atomic orbitals of the Mg $3p_z$, O $2p_z$, Ag $5s$, and Ag $5p_z$ states.
 The isosurfaces are shown for $\sim 60\%$ of their respective maximum values.  
}
\label{fig:orbitals}
\end{figure}

The next step in the analysis of the STM image formation is to begin further examination of the strength and nature of the Ag to MgO coupling across the interface.  First, we examine the relevant length scales by examining the sizes of the atomic orbitals in this system.  Fig.~\ref{fig:orbitals} shows isosurfaces of some of the important atomic orbitals on Ag, Mg, and O atoms.  While the O $2p$ orbitals are quite localized, the Ag and Mg orbitals are quite extended in space: although the interfacial O and Ag atoms are closest at the interface, the Ag to Mg tight-binding matrix elements must in fact be quite sizable as well given the extent of these atomic states; appreciable matrix elements can also be expected between the interfacial Ag and the top MgO layer.  These large sizes foreshadow the difficulties that will present themselves in any simple-minded analysis of the cross-interface coupling in terms of nearest neighbors and atomic arrangements localized at the interface.  A secondary implication is that although the LDOS at \EF isosurface plots in Figs.~\ref{fig:migs_3d}(a) and (b) seem to show only O $2p$ contributions in the MgO, this is in some ways deceptive as the compact O $2p$ orbitals have high probability densities that dominate the LDOS in real space.  The much more delocalized Mg orbitals may have significant weight but this is not visible in the plots unless one chooses extremely small isosurface values.  The importance of the Mg orbitals is discussed in more detail below.

We now switch to the L\"owdin representation of orthonormal atomic orbitals to aid in simplifying the analysis:  in the L\"owdin basis, the overlap matrices are identity by construction so we only have to consider the Hamiltonian matrix and its orthonormal eigenvectors.  We denote the Hamiltonian matrix at wave vector $\k$ in the L\"owdin representation as $\tilde H^\k_{\alpha\beta}$ which we decompose into sub-blocks corresponding to the Ag and MgO subsystems as
\begin{equation*}
\tilde H^\k = \left(
\begin{array}{cc}
\tilde H^\k_{mm} & \tilde H^\k_{mi} \\
(\tilde H^\k_{mi})^\dag & \tilde H^\k_{ii}
\end{array}
\right),
\end{equation*}
where $mm$ labels the metallic Ag subsystem, $ii$ labels the insulating MgO subsystem, and $mi$ labels the metal-to-insulator coupling elements.  The metallic subsystem Hamiltonian $\tilde H^\k_{mm}$ has states at \EF while the insulating one $\tilde H^\k_{ii}$ has an energy gap at \EF:  the coupling $H^\k_{mi}$ is responsible for creating evanescent states at energy \EF that propagate in the insulator.

Armed with this representation, we can first examine the nature of the states at \EF that propagate in the MgO.  For the subspace of MgO orbitals, the eigenstates of $\tilde H^\k_{ii}$ form a complete basis of valence and conduction bands, so we may ask how the density of states at \EF is decomposed into these two subspaces.  Letting $P_v^\k$ and $P_c^\k$ be orthogonal projectors onto the valence and conduction bands of the insulating MgO subsystem at wave vector $\k$, respectively, we compute the projected densities of states at \EF onto the MgO bands via
\[
L_v(E_{\mathrm{F}}) = \frac1{N_k}\sum_{n,\k} w(E_{\mathrm{F}}-E_{nk}) \bra{\psi_{n\k}}P_v^\k\ket{\psi_{n\k}}
\]
and
\[
L_c(E_{\mathrm{F}}) = \frac1{N_k}\sum_{n,\k} w(E_{\mathrm{F}}-E_{nk}) \bra{\psi_{n\k}}P_c^\k\ket{\psi_{n\k}}\,.
\]
The total density of states at \EF is
\[
D(E_{\mathrm{F}}) = \frac1{N_k}\sum_{n,\k} w(E_{\mathrm{F}}-E_{nk}) \braket{\psi_{n\k}}{\psi_{n\k}}\,.
\]
The second row of Table~\ref{tab:LvLc} shows these quantities for the relaxed 2ML MgO/Ag system ($d=2.68$) \AA).  We see projections for the states at \EF on both valence and conduction bands, and in fact the weight on the Mg-dominated conduction bands is higher than that of the O-dominated valence bands.  These results show that there are at least two independent, and in fact interfering (see below), paths for electronic wave functions to reach the surface.  In terms of comparing to computed STM images, however, it is more helpful to consider the projections onto the atomic states on the surface atoms
\[
  L_{\alpha}(E_{\mathrm{F}}) = \frac1{N_k}\sum_{n,\k} w(E_{\mathrm{F}}-E_{nk}) \bra{\psi_{n\k}}P_{\alpha}^\k\ket{\psi_{n\k}},
\]
where $P^{\k}_{\alpha}$ is the projector onto the (L\"owdin) atomic orbital $\alpha$.  The second row of Table~\ref{tab:La} displays these values for the relaxed 2 ML MgO/Ag system for the two orbitals that we have found dominate the STM image: O $2p_z$ and Mg $3p_z$ of the surface layer.  Both orbitals have weights at \EF which correlates  with the band projections of Table~\ref{tab:LvLc} and the STM image of Fig.~\ref{fig:tb_stm}(a).

When the MgO/Ag separation is reduced to $d=2.47$ \AA, the resulting two sets of projections are displayed in the first rows of Tables~\ref{tab:LvLc} and \ref{tab:La}.  The main effect of reducing the separation is to increase the projection onto the conduction band at the expense of the valence band, which in turn significantly increases the projection on the surface Mg $3p_z$ orbital and greatly increases the intensity at the Mg in the STM image Fig.~\ref{fig:tb_vs_dft}(d) at the expense of the O site.  This behavior is linked to the enlarged matrix elements between the extended Ag and Mg orbitals upon reduction of their separation.

\begin{figure}
  \centering\includegraphics[width=250pt]{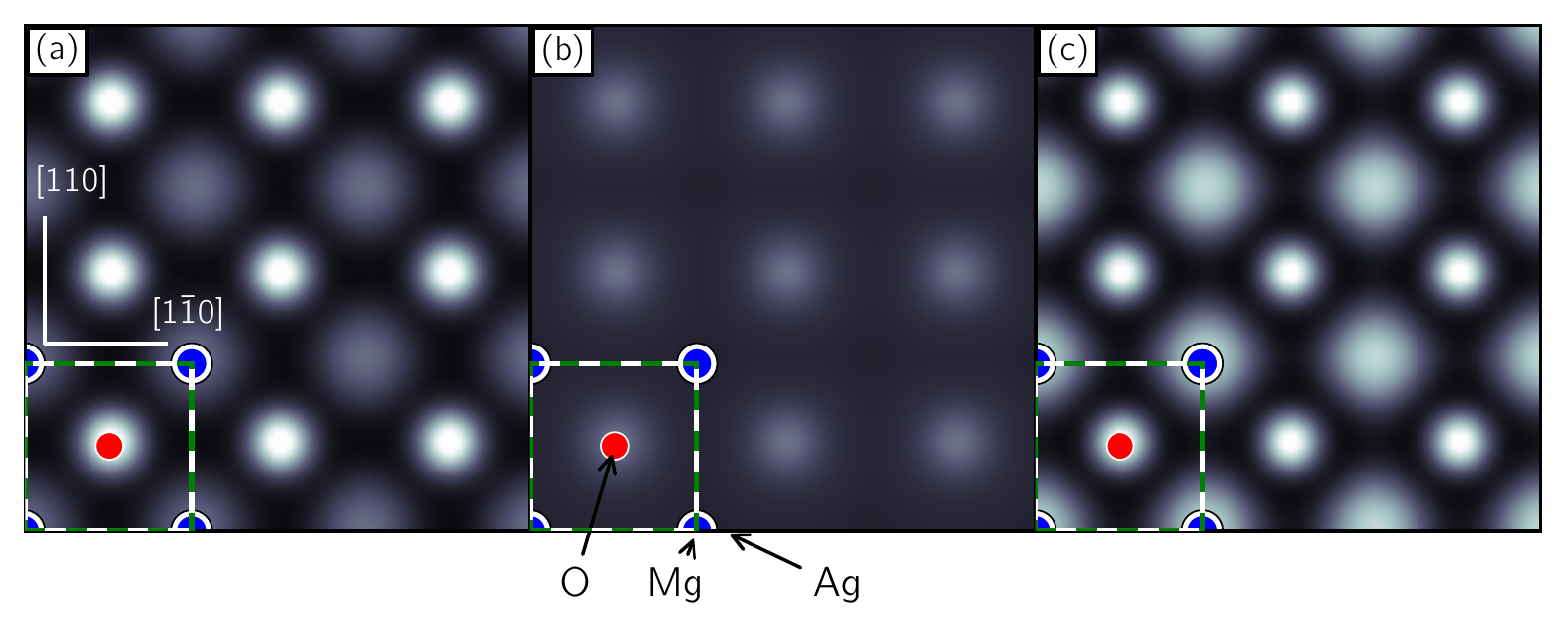}
\caption{
 Constant-height LDOS simulations of the 2 ML MgO/Ag(001) system
 based on the atomic-orbital tight-binding model and selective zeroing of the interfacial couplings $H^{\k}_{mi}$:
 (a) unperturbed Hamiltonian, 
 (b) zeroed Ag$_{\mathrm i}$ - Mg$_{\mathrm i}$ coupling,
 (c) zeroed Ag$_{\mathrm i}$ - O$_{\mathrm i}$ coupling.
 All images are computed at $1$~\AA\ above the surface.
 The blue (red) circles denote the positions of Mg (O) atoms on the surface layer.
  The white circles denote positions of Ag atoms
 at the interfacial layer.
 The dashed squares outline the unit cells used in the calculations.
}
\label{fig:tb_stm_zaps}
\end{figure}

A final set of manipulations on the system involves selectively removing certain Ag to MgO couplings and observing the result on the electronic structure at \EF  in the MgO.  The third rows of Tables~\ref{tab:LvLc} and \ref{tab:La} show the effect or removing (zeroing out) all entries in $H^\k_{mi}$ corresponding to orbitals on the interfacial silver atoms, denoted as Ag$_{\mathrm i}$, and all atomic orbital belong to the interfacial Mg, denoted as Mg$_{\mathrm i}$.  The resulting STM image is shown in Fig.~\ref{fig:tb_stm_zaps}(b).  Clearly, the interfacial Mg 
has a strong connection to the Ag as removing these connections makes the projections onto the valence and conduction bands of MgO drop by an oder of magnitude while reducing the LDOS at \EF on the surface orbitals as well.  This explains the generally dimming of the computed STM image.

However, zeroing the connection from Ag$_{\mathrm i}$ to all the orbitals of the interfacial oxygen O$_{\mathrm i}$ has a much more complex result.  Counter intuitively, the coupling to valence band actually {\it increases} compared to the pristine case, and the LDOS at \EF on the surface orbitals is greatly enhanced which correlates to STM image becoming brighter than before at both surface atomic sites as seen in Fig.~\ref{fig:tb_stm_zaps}(c).  When the coupling of Ag$_{\mathrm i}$ is zeroed to all orbitals of the interfacial layer (MgO)$_\mathrm{i}$, the projections onto the valence and conduction bands are reduced but are above their values when only Ag$_\mathrm{i}$-Mg$_\mathrm{i}$ couplings were removed, and the LDOS at \EF of the surface orbitals is greatly reduced for O $2p_z$ but quite large for Mg $3p_z$.  These results show interference:  zeroing an interfacial connection, Ag$_\mathrm{i}$-O$_\mathrm{i}$,  increases projections compared to the pristine case or when zeroing to the entire interfacial layer.  There are interfering paths for the propagation of the states at \EF across the interface determined by a complex interplay of Ag-O and Ag-Mg couplings.

We note that there is a complication in our analysis: the L\"owdin orthogonalization mixes atomic orbitals around neighboring sites that are coupled by overlap matrix elements, so that zeroing various $H^\k_{mi}$ entries is not an extremely spatially localized modification especially given the aforementioned large spatial extent of the Ag and Mg orbitals.  The zeroing procedure in the L\"owdin procedure is expected to be more insightful in systems with highly localized atomic orbitals.

\begin{table}
  \caption{\label{tab:LvLc}
Total density of states $D(E_{\mathrm{F}})$ and projections of the states at the Fermi level \EF onto the valence $L_v(E_{\mathrm{F}})$ and conduction bands $L_c(E_{\mathrm{F}})$ of the MgO film for the 2 ML MgO/Ag system using the tight-binding method.  The states at \EF were selected by a Fermi-Dirac smearing function with $kT=0.1$~eV. The $d=2.68$ \AA\ system is the fully relaxed interfacial system.  The $d=2.47$ \AA\ is described in the text and has a fixed and reduced MgO/Ag separation.  Starting at the third row, the dependence of densities of states and projections is shown when various interfacial matrix elements are set to zero.
}
\begin{ruledtabular}
\begin{tabular}{ccrrr}
System & Zeroed couplings & $D(E_{\mathrm{F}})$ & $L_v(E_{\mathrm{F}})$ & $L_c(E_{\mathrm{F}})$ \\
 &  & eV$^{-1}$ & eV$^{-1}$  & eV$^{-1}$\\
  \hline
 $d=2.47$ \AA & -- & 0.43 & 0.028  & 0.047\\
 $d=2.68$ \AA & -- & 0.44 & 0.029 & 0.041\\
              & & & & \\
 $d=2.68$ \AA & Ag$_{\mathrm i}$ - Mg$_{\mathrm i}$ & 0.24 & 0.0020 & 0.0033\\
 $d=2.68$ \AA & Ag$_{\mathrm i}$ - O$_{\mathrm i}$ & 0.46 & 0.051 & 0.046\\
 $d=2.68$ \AA & Ag$_{\mathrm i}$ - (MgO)$_{\mathrm i}$ & 0.25 & 0.0041 & 0.0035
\end{tabular}
\end{ruledtabular}
\end{table}

\begin{table}
  \caption{\label{tab:La}
    Density of states $L_{\alpha}(E_{\mathrm{F}})$ projected on the L\"owdin orbitals $\alpha$ of the surface atoms for the 2 ML MgO/Ag system using the tight-binding method.
The states at \EF were selected by a Fermi-Dirac smearing function with $kT=0.1$~eV. The $d=2.68$ \AA\ system is the fully relaxed interfacial system.
The $d=2.47$ \AA\ is described in the text and has a fixed and reduced MgO/Ag separation.
Starting at the third row, the dependence of densities of states and projections is shown when various interfacial matrix elements are set to zero.
}
\begin{ruledtabular}
\begin{tabular}{cccc}
  System & Zeroed couplings & $L_{\mathrm{O}\,2p_z}(E_{\mathrm{F}})$ &
  $L_{\mathrm{Mg}\,3p_z}(E_{\mathrm{F}})$ \\
         &  & $10^{-3}$eV$^{-1}$ & $10^{-3}$eV$^{-1}$  \\
  \hline
 $d=2.47$ \AA & -- & 1.13  & 0.66\\
 $d=2.68$ \AA & -- & 1.76  & 0.54\\
              & & & \\
 $d=2.68$ \AA & Ag$_{\mathrm i}$ - Mg$_{\mathrm i}$ & 0.36 & 0.53\\
 $d=2.68$ \AA & Ag$_{\mathrm i}$ - O$_{\mathrm i}$ & 3.73 & 1.54\\
 $d=2.68$ \AA & Ag$_{\mathrm i}$ - (MgO)$_{\mathrm i}$ & 0.07 & 1.15
\end{tabular}
\end{ruledtabular}
\end{table}
Summarizing the main findings of this section, the tight-binding approach is useful in that (a) it generates STM images which agree well with the full plane-wave results, (b) it allows one to unambiguously show that the STM image is generated by the electronic structure at \EF on the surface MgO layers, (c) that the MgO/Ag separation $d$ controls the relative importance of Ag-Mg coupling and coupling of the states at \EF to the conduction band of the MgO film, and (d) it illustrates the complex nature of the coupling across the interface and its relation to the formation of the STM image.  This coupling across this interfacial system is in fact quite delocalized in real space primarily due to the large spatial extent of the Mg orbitals in the MgO overlayer.  We expect similar complex and delocalized behavior at low bias in other metal/oxide interfaces where the key cationic atomic orbitals are large on the scale of the inter-atomic distances.  Conversely, if the cationic atomic states are localized to begin with, e.g., the $3d$ states of first row transition metals, the interpretation of the low-bias STM in terms of interfacial behavior may be significantly simplified.

\section{Summary}
\label{sec:summary}
We have investigated the STM contrast at near-zero-bias voltage
for thin MgO films on  Ag (001) substrates using DFT first-principles
simulations. We found that the STM images cannot be simply and directly attributed to the states
of the Ag substrates.  The STM image is in fact completely dominated by the contributions of the electronic states of the topmost MgO atomic plane on the surface of the film.  Hence, the STM image formation process is as follows:  metallic states at the Fermi level originate in the Ag, couple to the MgO atomic orbitals (or the MgO band states) across the interface, propagate through the insulating MgO lattice, and evanescently decay on their way to the surface.  The STM image thus is created by the amplitudes of these evanescent states at \EF on the surface atoms.  Our results show that the cross-interfacial coupling is complex and long-ranged in this system, defying the simplest nearest-neighbor analysis in terms of contributions solely at short range across the interface.  The complex behavior is caused primarily by the large spatial extent of the Mg $3s$, $3p$, and $3d$ states that dominate the conduction band of the MgO film.  We have observed that there are at least two paths for the propagation of the electronic states across the interface, and that they interfere in a complicated manner when forming the STM image above the surface.

In the process of the analysis, we have developed a simple tight-binding method that successfully reproduces the STM contrast computed from the more accurate plane-wave calculations.  The tight-binding approach permits a variety of analyses to be performed on how the STM image is formed, what information it carries, and the key atomic orbitals that determine its overall behavior.  The method is general and applicable to other interfacial systems.

Finally, while this particular interfacial system features delocalized couplings which make simple analysis difficult, the main culprit is the large extent of the cation Mg atomic states that dominate the conduction bands of the MgO.  Hence, systems where both the conduction and valence bands of the insulating overlayer are dominated by localized orbitals are preferred:  in such a situation, one has a better chance of extracting information on the localized behavior of the buried information from the STM image on the surface.  For example, metal oxide films incorporating $3d$ transition metals should be good candidates for future studies due to the spatial locality of the $3d$ orbitals. 

\acknowledgments

A. M. acknowledges useful discussions with Matthew S. J. Marshall.  This work was supported by NSF MRSEC DMR
1119826 and by the facilities
and staff of the Yale University Faculty of Arts and Sciences High
Performance Computing Center.  Additional computations used the NSF
XSEDE resources via grant TG-MCA08X007.


\begin{thebibliography}{30}
\bibitem [{exp({\natexlab{b}})}]{explan_02}%
  \BibitemOpen
  \href@noop {} {} \bibinfo {note} {If we allow structural
  relaxations the atomic positions in the MgO layer change at most by
  $0.1$~\AA\ and we see no significant changes in the STM images.}\BibitemShut
  {Stop}%
\end{thebibliography}
\end{document}